\documentclass[twocolumn,nofootinbib,preprintnumbers,amsmath,amssymb]{revtex4}
\usepackage{graphicx}
\usepackage[normalem]{ulem}
\usepackage{hyperref,color}
\newcommand{\uu}{{\bf u}}
\newcommand{\vv}{{\bf v}}

\newcommand{\im}{\mbox{Im}}
\newcommand{\be}{\begin{equation}}
\newcommand{\ee}{\end{equation}}
\newcommand{\ba}{\begin{eqnarray}}
\newcommand{\ea}{\end{eqnarray}}
\newcommand{\new}{\textcolor{black}}
\begin{document}
\title{Comment on ``Explaining the specific heat of liquids based on instantaneous normal modes''}

\author{Walter Schirmacher}
\affiliation{Institut f\"ur Physik, Universit\"at Mainz, D-55099 Mainz, Germany}
\affiliation{Center for Life Nano Science @Sapienza, Istituto Italiano di Tecnologia, 295 Viale Regina Elena, I-00161, Roma, Italy}

\author{Taras Bryk}
\affiliation{Institute for Condensed Matter Physics of the National Academy of Sciences of Ukraine, UA-79011 Lviv, Ukraine}
\affiliation{Institute of Applied Mathematics and Fundamental Sciences, Lviv National Polytechnic University, UA-79013 Lviv, Ukraine}

\author{Giancarlo Ruocco}
\affiliation{Center for Life Nano Science @Sapienza, Istituto Italiano di Tecnologia, 295 Viale Regina Elena, I-00161, Roma, Italy}
\affiliation{Dipartimento di Fisica, Universita' di Roma ``La Sapienza'', I-00185, Roma, Italy}
\begin{abstract}
	In a recent paper (Phys. Rev. E {\bf 104}, 014103 (2021) ) M. Baggioli
	and A. Zaccone formulate a theoretical description of 
the specific heat of liquids
	by using Debye's expression for the specific heat of solids
	and inserting a density of states (DOS) which they claim to represent
	the instantaneous-normal-mode (INM) spectrum of a liquid. However,
	the quantum-mechanical procedure of Debye cannot be used
	for a classical liquid and the authors' formula for the INM spectrum
	does not represent the known INM spectra of simple liquids. Furthermore, 
	the derivation of this formula from their model equation of motion
        is mathematically in error. 
	Finally experimental test of the
	theory for the specific heat
	of {\it liquids} is performed by
	fitting the data of {\it supercritical fluids}.
	\new{To our opinion,}
	these
	and a lot of other inconsistencies render this
        work not suitable for studying the specific
	heat of liquids.
\end{abstract}

\maketitle

Baggioli and Zaccone (BZ)
present in a recent paper
\cite{baggioli21} 
a treatise on the specific heat of liquids. They claim
that this would be a fundamental new theory for the specific
heat of liquids. 
\new{The authors end with the statement:
``To summarize, the above theory provides a definitive
answer to the mystery of liquid specific heat and ideally
completes the agenda of the kinetic theory of matter, set
over 100 yr ago by Debye, Einstein, Planck, and co-workers.''}

By contrast, \new{on the one hand, the specific heat of liquids
is not a mystery, on the other hand,}
the paper turns out to contain so many elementary scientific
and mathematical inconsistencies and 
errors, that it does not
appear to be of any value for studying the specific heat of liquids.

BZ ($i$)
formulate a Debye-like quantum theory,
in which they treat the instantaneous-normal (INM) modes, which
are the eigenmodes of the Hessian of a simulated classical
liquid \cite{stratt95,keyes97} as bosons in an interacting
quantum liquid, referring, among others,
to papers on quantum-chromodynamics
\cite{sirlin93}. They repeatedly 
claim that there would be no satisfactory
theory for the specific heat of liquid available yet,
ignoring the existing literature on the thermodynamics
of simple liquids \cite{hansenmcdonald,rosenfeld98,dyre13}.
What BZ have done, boils down to ($ii$) inserting
an expression for the density of states for
the INM modes, derived by BZ previously \cite{baggioli21a},
into Debye's expression
for the quantum specific heat of a solid. 
They ($iii$) claim
to have good agreement with specific heat data on simple liquids.

($i$) It is well known that quantum effects are not relevant to
most liquids (the exceptions are the helium liquids), because
in the liquid state the thermal de-Broglie wavelength
$\lambda=h/[mk_BT]^{1/2}$ is much smaller than
the diameter of the liquid particles. Here
$h$ and $k_B$ are Planck's and Boltzmann's constants, 
$m$ is the particle mass, and $T$ is the temperature.

\new{BZ motivate their quantum approach by discussing}
a weakly interacting Bose
gas with elementary excitations of energy $\hbar|\omega_q|$,
where they obviously identify the frequencies $\omega_q$
with those appearing in Eq. (\ref{inm}) (see below), the
modulus of the square-root of the eigenvalues of the
instantaneous
Hessian matrix of a {\it classical} liquid. The dispersion
$\omega_q$ is not specified. As mentioned above, BZ
invoke papers on unstable (massive) bosons in quantum
chromodynamics as justification of their identification.

For classical liquids and gases, the partition function
factorizes into a factor arising from integrating over the
kinetic energy and the configurational factor arising from
integrating over
the potential energy (configuration integral). 
This means (as is well known \cite{hansenmcdonald}), that
the dynamics of a classical liquid does not enter
into its thermodynamic properties.

For simple liquids in an $(N,V,T)$ system, the
energy equation of states is given by
the sum of the ideal and the excess term
\ba
\frac{1}{N}U(N,V,T)&=&
\frac{1}{N}\bigg(U^{id}+U^{ex}\bigg)\\
&=&\frac{3}{2}k_BT+2\pi\rho(T)\int d^3{\bf r}\phi(r)g(r,T)\,\nonumber 
\ea
and the specific heat per particle 
$c_V(T)=c_V^{id}+c_V^{ex}$
is just the derivative with respect
to the temperature $T$.
Here $\phi(r)$ is the parwise potential and $g(r,T)$ the radial pair distribution
function, for which well-established thermodynamic theories exist,
notably thermodynamic perturbation theory \cite{hansenmcdonald}. For example,
Rosenfeld and Tarazona \cite{rosenfeld98} (not cited in \cite{baggioli21})
use density-functional theory and thermodynamic perturbation theory to
come up with an expression of the excess internal energy $U^{ex}$ and the corresponding  specific heat $c_V^{ex}$, which depend on the temperature
$T$ via a power law $U^{ex}\propto T^{3/5}$, $c_V^{ex}\propto T^{-2/5}$
, which stems from the singularity induced by the presence
of the atomic hard cores. 
It has been demonstrated that this equation of state and
the corresponding free-energy functional describe
the thermodynamics of liquids rather well,
in partcular the temperature dependence of the
specific heat
\cite{rosenfeld98,dyre13}.

Conclusion: At variance with the claims of BZ, a  quantum description of simple liquid is
not adequate, and a well-established thermodynamic theory for classical liquids,
including the specific heat, is available. 

\begin{figure}
\includegraphics[width=8.5cm]{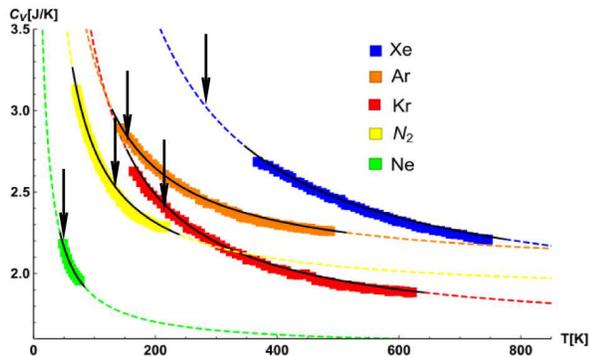}
	\caption{Specific heat data for some fluid elementary
	materials, fitted by Baggioli and Zaccone \cite{baggioli21}
	with the DOS of Eq. (..) inserted into the Debye formula.
	The arrows indicate the liquid-gas critical temparatures
	$T_c$: Ne (44 K), Ar (150 K), Kr (210 K), Xe (290 K), N$_2$ (126.2 K).
		}\label{bz-nob1}
\end{figure}

($ii$) We turn now to the BZ formula for the
density of states of the instantaneus-normal
modes of a simulated model for a classical liquid.
In the INM literature it had been customary to represent the
INM spectrum, i.e. the normalized histogram $\rho(\lambda)$ of eigenvalues 
$\lambda_i=\omega_i^2$
of the Hessian matrix of the potential energy of a liquid at a certain
time instant, as
\be\label{inm}
g(\omega)=\left\{\begin{array}{ccc}
2\omega\rho(\lambda)&for&\lambda=\omega^2\geq 0)\\
	&&\\
2|\omega|\rho(\lambda)&for&\lambda=\omega^2< 0)\, ,\\
\end{array}\right.
\ee
where the absolute sign $|\dots|$ refers to that of a complex number, and the unstable part of the spectrum is plotted
along the negative $\omega$ axis.
At very high frequencies $g(\omega)$ can be taken to
represent something like the density of states
for vibrations, which are known to exist at such frequencies.
At low and negative values of $\lambda$ the modes represent
unstable configurations.
BZ ackowledge this
and design a 
model 
for the DOS, motivated by a Langevin-type equation
(see below)
\cite{baggioli21a,baggioli21}
\be\label{bz1}
g(\omega)\propto\sum_i\frac{\omega\Gamma_i}{\omega^2+\Gamma_i^2}
\ee
According to this equation
BZ \cite{baggioli21a} claimed that the function $g(|\omega|)$ would
be proportional to $|\omega|$ for small $\lambda$. This would mean that
$\rho(\lambda)$ would be constant in this regime. However it 
is known since long time
\cite{sastry01,taraskin02a} and has recently been explained
\cite{schirm22}, that
$\rho(\lambda)$ is strongly peaked at small values of $\lambda$, i.e. far from being
constant. $\rho(\lambda)$ is also not universal, as claimed
by BZ \cite{baggioli21a},
but depends
strongly on temperature \cite{schirm22}.
Furthermore, BZ claim to have derived expression (\ref{bz1}) from a Langevin
equation (strange enough without fluctuating forces) for the local velocities $\vv_i$ with
damping coefficients $\Gamma_i$

\be\label{bz2}
\frac{d}{dt}\vv_i=-\Gamma_i\vv_i
\ee
The vibrational DOS, however, has to be calculated from the
imaginary part of the Green's function for the \mbox{\it displacements}
$\uu_i$ with $\vv_i=\frac{d}{dt}\uu_i$. For model (\ref{bz2}) this
Green's function takes the form
\be
G_{ii}(\omega)=\frac{1}{-i\omega}\frac{1}{-i\omega+\Gamma_i}
\ee
Taking the imaginary part we arrive at
\ba
g(\omega)&\propto&\sum_i\,2\omega\im\{G_{ii}(\omega)\}\\
&\propto& \sum_i\frac{\Gamma_i}{\omega^2+\Gamma_i^2}\, ,\nonumber
\ea
which is at variance with BZ' expression (\ref{bz1}), where
obviously a factor $1/\omega$ is missing.
Certainly this,
together with the inadequateness of (\ref{bz1}) to describe
simulated INM spectra \cite{schirm22},
invalidates all reasoning in Refs. 
\cite{baggioli21} and \cite{baggioli21a}.

($iii$) Let us now look at the comparison with experiment. BZ
modified their formula of Ref. \cite{baggioli21a} as follows:
\be\label{bz3}
g(\omega)\propto\frac{\omega}{\omega^2+\Gamma^2}e^{-\omega^2/\omega_D^2}\, ,
\ee
replacing Debye's cutoff with a soft Gaussian.
They assumed that the damping constant
obeys an Arrhenius temperature law $\Gamma=\Gamma_0 e^{-\epsilon/k_BT}$, where $\epsilon$ is the Lennard-Jones depth parameter and $\Gamma_0$ is a prefactor.

Fits with inserting $g(\omega)$
according to Eq. (\ref{bz3})
into Debye's formula
for the specific heat have been performed for
the inert gases neon, argon, krypton and xenon, as well
as nitrogen.
BZ took values for the Debye frequency from literature
of solid rare gases and $N_2$, together with known values for
the Lennard-Jones parameter $\epsilon$ and took the
prefactor $\Gamma_0$ as fit parameter.

In Fig. 1 we report the data from Fig. 2 of BZ,
and we add arrows indicating 
the appropriate values of the critical temperatures
for the investigated materials.
We observe that the data fitted by BZ with
formula (\ref{bz3}) are predominantly 
in the {\it supercritical} states, i.e. the 
liquid theory has been tested against supercritical fluids.
So BZ use a quantum theory
and values of the Debye frequency of solids to
fit supercritical gases, not liquids, and take this
as evidence for the descriptive power of a theory
for a liquid.

There are more elementary inconsistencies in Ref. \cite{baggioli21}
, e.g. calling
the specific heat law for a free one-atomic
gas $c_V^{id}=\frac{3}{2}k_B$ the {\it Dulong-petit law}. 
As is well known, the
latter holds
for a solid and gives the double value due to the 
additional presence of the
harmonic vibrational degrees of freedom.

We conclude that both publications of BZ, 
Ref. \cite{baggioli21} and Ref. \cite{baggioli21a}
are not helpful for understanding the dynamics and
thermodynamics of simple liquids. 

\subsection*{Acknowledgment}
TB was supported by NRFU grant 2020.02/0115.

\end{document}